\documentclass[prl,twocolumn,superscriptaddress,showpacs]{revtex4}
\usepackage[dvips]{graphicx}
\usepackage{amsmath,amssymb}

\begin{document}
\title{Simple and efficient quantum key distribution with parametric down-conversion}

\author{Yoritoshi Adachi}
\email{adachi@qi.mp.es.osaka-u.ac.jp}
\affiliation{Division of Materials Physics, Department of Materials
Engineering Science, Graduate School of Engineering Science, Osaka
University, Toyonaka, Osaka 560-8531, Japan}
\affiliation{CREST Research Team for Photonic Quantum Information, 4-1-8
Honmachi, Kawaguchi, Saitama 331-0012, Japan}

\author{Takashi Yamamoto}
\affiliation{Division of Materials Physics, Department of Materials
Engineering Science, Graduate School of Engineering Science, Osaka
University, Toyonaka, Osaka 560-8531, Japan}
\affiliation{CREST Research Team for Photonic Quantum Information, 4-1-8
Honmachi, Kawaguchi, Saitama 331-0012, Japan}

\author{Masato Koashi}
\affiliation{Division of Materials Physics, Department of Materials
Engineering Science, Graduate School of Engineering Science, Osaka
University, Toyonaka, Osaka 560-8531, Japan}
\affiliation{CREST Research Team for Photonic Quantum Information, 4-1-8
Honmachi, Kawaguchi, Saitama 331-0012, Japan}

\author{Nobuyuki Imoto}
\affiliation{Division of Materials Physics, Department of Materials
Engineering Science, Graduate School of Engineering Science, Osaka
University, Toyonaka, Osaka 560-8531, Japan}
\affiliation{CREST Research Team for Photonic Quantum Information, 4-1-8
Honmachi, Kawaguchi, Saitama 331-0012, Japan}
\date{\today}
\begin{abstract}
We propose an efficient quantum key distribution protocol based on the
 photon-pair generation from parametric down-conversion (PDC). 
It uses the same experimental setup as the conventional protocol, but a
 refined data analysis enables detection of photon-number splitting
 attacks by utilizing information from a built-in decoy state.
Assuming the use of practical detectors, we analyze the unconditional 
security of the new scheme and show that it
improves the secure key generation rate by several orders of magnitude
at long distances, using a high intensity PDC source.
\end{abstract}

\maketitle
Quantum key distribution (QKD) is a promising application of quantum
information, with which two distant legitimate
users (the sender Alice and the receiver Bob) can share a common random
bit string, known as a secret key, with negligible leak to an
eavesdropper Eve. 
The first QKD protocol has been proposed by Bennett and Brassard in 1984, which is called BB84 \cite{BB84}.
The original BB84 protocol proposes the use of an ideal single-photon
source, and secure key distribution should be possible up to the distance
at which Bob's photon detection rate and his dark counting rate are 
comparable. Since such an ideal single-photon source is not available 
today, weak coherent pulses (WCPs) from attenuated lasers are
commonly used as a photon source \cite{GYSapl04,Stucki02,Kimura04,Zhao06PRL06,Peng06qph,Hiskett06qph}. 
The WCP has two imperfections, the multi-photon part and 
the vacuum part. The multi-photon part is vulnerable against 
photon-number splitting (PNS) attacks \cite{BLMSprl00}, and one must reduce 
the energy of the WCP in order to reduce the fraction of 
the multi-photon part. This leads to a very low key rate.
The existence of the vacuum part simply leads to a reduction
of Bob's photon detection rate, resulting in a shorter distance 
limit. Recent analyses \cite{Wangprl05,LMCprl05} show that 
the former problem can be avoided by randomly mixing pulses 
with different energies (decoy states) \cite{Hwangprl03}. 
But about half of the pulses are still in the vacuum state, 
and hence the distance limit falls short of the one with
the ideal single-photon source. 

\begin{figure}[t]
\begin{center}
\includegraphics[scale=1]{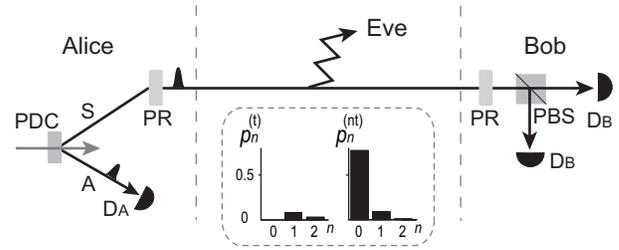}
\caption{
The experimental setup of QKD system with PDC.
Alice and Bob choose the bases by polarization rotators (PR's). 
Bob detects the photons by two threshold detectors (D$_{\rm B}$'s) after a polarizing beam splitter (PBS). 
The inset shows 
the photon number distributions of the triggered events $p_n^{\rm (t)}$ and the nontriggered events $p_n^{\rm (nt)}$, when $\mu=0.3$, $\eta_{\rm A}=0.5$ and $d_{\rm A}=10^{-6}$. 
}
\label{fig:setup}
\end{center}
\end{figure}

Another candidate of photon sources within reach of current
technology is conditional generation of single photons 
based on parametric down-conversion (PDC) \cite{Lutkenhauspra00}.
The state of the photons generated in two modes ${\rm A}$ and
${\rm S}$ by PDC can be written as \cite{YPpra87}
\begin{eqnarray}
 |\Psi\rangle_{\rm AS} &=&
\sum ^{\infty}_{n=0} \sqrt{\mathstrut p_n}
|n\rangle_{\rm A}|n\rangle_{\rm S}, 
\label{eq:PDC} \\
p_n&\equiv& \mu^n(1+\mu)^{-(n+1)}
\label{eq:PDC2}
\end{eqnarray}
where $|n\rangle$ represents the state of $n$ photons 
and $\mu$ is the average photon-pair rate. 
If Alice measures the mode ${\rm A}$ by an ideal photon-number-resolving
detector with unit efficiency and selects the cases where 
just one photon has been detected, she would conditionally obtain 
an ideal single photon in mode ${\rm S}$. But in practice, she must 
use a threshold (on/off) detector with nonunit efficiency, which cannot
distinguish one from two or more photons. In this case, 
she selects the cases where the detection has occurred (triggered events).
The good news is 
that the dark count rate of current detectors is very low, 
and we can still neglect the vacuum part of mode ${\rm S}$ for 
triggered events (see Fig.~\ref{fig:setup}).
Hence this source achieves the same distance limit as the 
ideal source. On the other hand, the mode ${\rm S}$ contains 
multi-photons, which is the same drawback as the WCP.
One must decrease $\mu$ and thereby reduce the rate of triggering 
to avoid PNS attacks, leading to a severely low key rate.
The remedies for this problem proposed so far are accompanied by  
introduction of additional complexity to the experimental setup,
such as the random amplitude modulation for the use of decoy states
and/or replacing Alice's detector by detector arrays in space or in 
time domain to improve the photon-number-resolving ability \cite{HKpra06,MS06}.

In this letter, we propose a very simple solution. Nothing is added 
to the experimental setup of the PDC with a triggering detector. 
The crux of our new protocol is to run 
the BB84 protocol regardless of whether Alice's detector is 
triggered or not. By comparing the detection rates 
for the triggered events and the nontriggered events, we can detect  
the presence of PNS attacks. We assume that threshold detectors
are used by Alice and Bob, and derive a formula for the unconditionally
secure key rate. Borrowing the parameters in a recent experiment,
our calculation shows that the key rate is improved by 
several orders of magnitude compared to the conventional 
security analysis.

We first look at the property of Alice's source composed of 
PDC with Eq.~(\ref{eq:PDC}) and a threshold detector 
D$_{\rm A}$ with efficiency $\eta_{\rm A}$ and dark count rate $d_{\rm A}$.
Let $\gamma_{n}$ be the probability of 
detection (triggering) at D$_{\rm A}$ when 
$n$ photons are emitted in mode ${\rm S}$.
Since $n$ photons are emitted also in mode A, we have
\begin{eqnarray}
\gamma_{n}=1-(1-d_{\rm A})(1-\eta_{\rm A})^n. 
\label{eq:POVMt}
\end{eqnarray}
Then  
$n$-photon emission events (at rate $p_n$) are divided into 
the events with triggering (at $p_n^{\rm (t)}$)
and the events without triggering (at $p_n^{\rm (nt)}$),
where 
$
p_n^{\rm (t)}= p_n \gamma_{n}
$
and 
$
p_n^{\rm (nt)}= p_n (1-\gamma_{n}), 
$
whose distributions typically look like Fig.~\ref{fig:setup}.

Alice changes the polarization of the pulse in 
mode ${\rm S}$ according to 
the BB84 protocol and sends it to Bob. Bob measures this signal 
by a polarization rotator and a polarizing beam splitter 
followed by two threshold detectors, as in Fig.~\ref{fig:setup}.
We say the signal is `detected' by Bob if at least one 
of the detectors clicks. When both detectors click, Bob
assumes his outcome to be a random bit value.
Let $Q_n$ be the rate of events where 
Alice emits $n$ photons in mode ${\rm S}$ and Bob detects the signal.
These events are also divided into two groups, 
the events accompanied by Alice's triggering (at rate $Q_n^{(\rm t)}$)
and the rest (at $Q_n^{(\rm nt)}$), where 
$
Q_n^{\rm (t)}= Q_n \gamma_{n}
$
and 
$
Q_n^{\rm (nt)}= Q_n (1-\gamma_{n}).
$
Behind these relations lies the fact that the state of 
PDC in Eq.~(\ref{eq:PDC}) becomes a direct product once 
we condition on the photon number $n$ in mode ${\rm S}$.
Hence there should be no correlations between 
the triggering at D$_{\rm A}$ and any event
occurring in mode ${\rm S}$. This fact also ensures that
the quantum bit error rate (QBER) $e_n$ when Alice emits 
$n$ photons in mode ${\rm S}$ should be 
the same whether or not the triggering occurs at D$_{\rm A}$.
Therefore, the overall detection rate $Q^{\rm (t)}$ and 
the QBER $E^{\rm (t)}$ with triggering, and 
the overall detection rate $Q^{\rm (nt)}$ and 
the QBER $E^{\rm (nt)}$ without triggering are expressed 
by
\begin{eqnarray}
Q^{\rm (t)}=\sum^{\infty}_{n=0}Q^{\rm (t)}_n,\quad
Q^{\rm (nt)}=\sum^{\infty}_{n=0}Q^{\rm (nt)}_n,
\label{eq:Qsum}
\\
E^{\rm (t)}=\sum^{\infty}_{n=0}\frac{Q^{\rm (t)}_n e_n}{Q^{\rm (t)}}, \quad
E^{\rm (nt)}=\sum^{\infty}_{n=0}\frac{Q^{\rm (nt)}_n e_n}{Q^{\rm
 (nt)}}.
\label{eq:Esum}
\end{eqnarray}
These four quantities are observed in the actual protocol,
while there is no way to measure directly 
the contributions from each photon number, except for $e_0$,
which is always $1/2$.

We discuss the security of our protocol 
by Gottesman-Lo-L\"utkenhaus-Preskill formula \cite{GLLP04,Lo05}, which is
written as follows for the key rate $R^{\rm (t)}$
with triggering:
\begin{eqnarray}
R^{\rm (t)}&=&q\{-Q^{\rm (t)}f(E^{\rm (t)})H_2(E^{\rm (t)})\nonumber\\
&&+Q^{\rm (t)}_0+Q^{\rm (t)}_1[1-H_2(e_1)]\}.
\label{eq:MGLLP}
\end{eqnarray}
The formula has recently been proved \cite{Koashi06}
to be valid even if Bob's detection is made by threshold detectors
as in Fig.~\ref{fig:setup}, as long as the two detectors have the same efficiency.
Here $q(=1/2)$ is the protocol efficiency, 
$f(E)$ is the error correction efficiency, 
and $H_2(E)$ is the binary entropy function. 
Since $Q^{\rm (t)}_0$, $Q^{\rm (t)}_1$, and $e_1$ are not 
exactly determined in the actual protocol, we must adopt 
the worst value of $R^{\rm (t)}$ in the possible range of 
these parameters.

\begin{figure}[t]
\begin{center}
\includegraphics[scale=1]{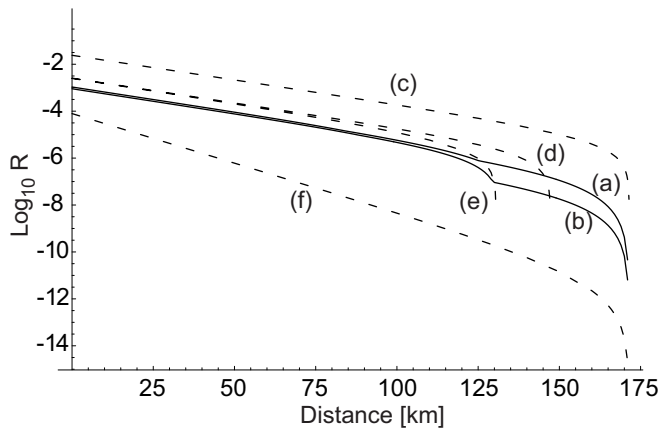}
\caption{
Achievable key rates for different implementations of BB84.
The calculations are done in the case of 
 (a) the efficient PDC protocol with $\eta_{\rm A}=0.5$ and
 $d_{\rm A}=10^{-6}$,
(b) the efficient PDC protocol with $\eta_{\rm A}=0.1$ and
 $d_{\rm A}=10^{-6}$,
(c) ideal single-photon source,  
(d) WCP with infinite number of decoy states,  
(e) WCP with one decoy state, and 
(f) the conventional PDC protocol with $\eta_{\rm A}=1$ and $d_{\rm A}=0$.  
}
\label{fig:graph}
\end{center}
\end{figure}

In the conventional protocol, we only observe $Q^{\rm (t)}$ and 
$E^{\rm (t)}$. In this case, we rely on the obvious inequality 
$Q^{\rm (t)}_n\le p^{\rm (t)}_n$ to obtain an upper bound on
the multi-photon contribution 
$Q^{\rm (t)}_{\rm multi}\equiv \sum^{\infty}_{n=2}Q^{\rm (t)}_n$.
This bound is meaningful only when $Q^{\rm (t)}> p^{\rm (t)}_{\rm multi}\equiv 
\sum^{\infty}_{n=2} p^{\rm (t)}_n$.
Since the scaling to $\mu$ and the channel transmission $\eta_{\rm c}$
is $Q^{\rm (t)}\sim O(\eta_{\rm c} \mu)$ and 
$p^{\rm (t)}_{\rm multi}\sim O(\mu^2)$, we have to choose $\mu\sim O(\eta_{\rm c})$
and hence $R^{\rm (t)}\sim O(\eta_{\rm c}^2)$ at best, which means 
a rapid decrease of the key rate against the distance [see Fig.~\ref{fig:graph} (f) below].

Now we will show that observation of nontriggered events, 
$Q^{\rm (nt)}$ and $E^{\rm (nt)}$, leads to a significant 
improvement of the key rate. The crucial relation is 
\begin{eqnarray}
Q^{\rm (t)}_n=r_nQ_n^{\rm (nt)}, 
\label{eq:relation}
\end{eqnarray}
where $r_n\equiv \gamma_n/(1-\gamma_n)= p^{\rm (t)}_n/p^{\rm (nt)}_n$.
 Eve cannot alter $r_n$ since it is determined by Alice's parameters $\eta_{\rm A}$ and
$d_{\rm A}$. 
From Eq.~(\ref{eq:POVMt}), we see
\begin{eqnarray}
0\leq r_0<r_1<r_2<r_3\cdots.
\label{eq:order}
\end{eqnarray}
By comparing $r\equiv Q^{\rm (t)}/Q^{\rm (nt)}$ with $r_n$'s,
we have a clue about the distribution $Q^{\rm (t)}_n$
over the photon number. The mechanism can be explained 
in two different ways. If we assume Alice's measurement 
by D$_{\rm A}$ occurs earlier, then it looks as if 
she randomly switches between two distributions, $\{p^{\rm (t)}_n\}$
and $\{p^{\rm (nt)}_n\}$. 
This is rather similar to the idea of one-decoy-state QKD \cite{LMCprl05}.
Comparing $r$ and $r_n=p^{\rm (t)}_n/p^{\rm (nt)}_n$
gives a clue about the PNS attacks, namely, $r$ should be close to
$r_1$ in the normal operation, but it will approach $r_2$ if Eve
exploits the multi-photon events.
If we assume Alice's measurement occurs after Bob's detection,
we notice that the photon number distribution at mode A 
conditioned on Bob's detection is proportional to $Q_n$.
Hence Alice physically possesses the distribution about which 
she wants to learn, and she makes a measurement by D$_{\rm A}$.
The averaged rate $Q^{\rm (t)}/(Q^{\rm (nt)}+Q^{\rm (t)})$ should 
then be compared with $\gamma_n$, which is equivalent to the comparison
between $r$ and $r_n=\gamma_n/(1-\gamma_n)$.

The remaining question is whether such a clue is enough to 
improve the key rate significantly. In the decoy state methods,
we can tailor the number and the amplitudes of decoy states 
at will, but here we have no such freedom except for 
the strength $\mu$ of PDC. This is answered by 
conducting a quantitative analysis as follows.
From Eqs.~(\ref{eq:relation}) and (\ref{eq:order}),
we have 
$r_2Q^{\rm (nt)}_n \leq Q_n^{\rm (t)}$ for $n\ge 2$.
 Applying Eq.~(\ref{eq:Qsum}) leads to   
$r_2(Q^{\rm (nt)}-Q_0^{\rm (nt)}-Q_1^{\rm (nt)})\leq Q^{\rm
(t)}-Q_0^{\rm (t)}-Q_1^{\rm (t)}=rQ^{\rm (nt)}-r_0Q_0^{\rm
(nt)}-r_1Q_1^{\rm (nt)}$.
We thus obtain
the minimum value of $Q_1^{\rm (nt)}$ as a function of 
the only remaining unknown parameter $x\equiv Q_0^{\rm (nt)}/Q^{\rm (nt)}$:
\begin{eqnarray}
\frac{Q_1^{\rm (nt)}}
{Q^{\rm (nt)}}
\geq\frac{r_2-r-(r_2-r_0)x}{r_2-r_1}
\equiv \xi(x)
\label{eq:newQntL}.
\end{eqnarray}
From Eqs.~(\ref{eq:Esum}) and (\ref{eq:relation}) with $e_0=1/2$,
an upper bound on $e_1$ is given by 
\begin{eqnarray}
e_1&\leq& [Q^{\rm (t)}E^{\rm (t)}-Q_0^{\rm (t)}e_0]/Q_1^{\rm (t)}
\nonumber \\
&\leq&\frac{2rE^{\rm (t)}-r_0x}{2r_1\xi(x)}
\equiv \epsilon_{\rm t}(x).
\label{eq:e1t}
\end{eqnarray}
In a similar way, we have another bound
\begin{eqnarray}
e_1
\leq\frac{2E^{\rm (nt)}-x}{2\xi(x)}
\equiv \epsilon_{\rm nt}(x).
\label{eq:e1nt}
\end{eqnarray} 
Combining the two bounds, we have
\begin{eqnarray}
e_1\le \epsilon(x)\equiv
\min\{\epsilon_{\rm t}(x) ,\epsilon_{\rm nt}(x) \}
\label{eq:newe1U}.
\end{eqnarray}
Consequently, in the limit of large block size with which 
the estimation errors are negligible,
the key rate from the triggered events is given by 
\begin{eqnarray}
&&R^{\rm (t)}/q=
-Q^{\rm (t)}f(E^{\rm (t)})H_2(E^{\rm (t)})
\nonumber\\
&&+Q^{\rm (nt)}\min_x \{r_0 x 
+r_1 \xi(x)[1-H_2(\epsilon(x))]\},
\label{eq:MGLLPnew}
\end{eqnarray} 
where the minimum is taken over the range
$0\leq x \leq {\rm min}\{2 E^{\rm
(t)}(r/r_0),2E^{\rm (nt)}\}$.
This minimization should be  
numerically calculated in general, and we give examples later.
Before that, we here discuss the scaling of the key rate 
$R^{\rm (t)}$ against the channel transmission $\eta_{\rm c}$.
Up to the distance at which the influence of 
the dark countings of Bob's detectors becomes substantial, 
the error rates $E^{\rm (t)}$ and $E^{\rm (nt)}$ are 
almost independent of $\eta_{\rm c}$. The detection 
rates $Q^{\rm (t)}$ and $Q^{\rm (nt)}$ are both proportional 
to $\eta_{\rm c}$, and their ratio $r$ is also 
independent of $\eta_{\rm c}$. Then, the functions
$\xi(x)$, $\epsilon_{\rm t}(x)$, and $\epsilon_{\rm nt}(x)$
are independent of $\eta_{\rm c}$, and hence 
the key rate in Eq.~(\ref{eq:MGLLPnew}) scales as 
$R^{\rm (t)}\sim O(\eta_{\rm c})$. 
The PDC strength $\mu$ only affects the constant factor here, and
its optimum value is independent of $\eta_c$.
This is a significant 
improvement over the rate of the conventional protocol, 
$R^{\rm (t)}\sim O(\eta_{\rm c}^2)$.

When the distance is not so large, we may produce a secret key 
also from the nontriggered events. In this case, it is more 
efficient when the error reconciliation is separately applied 
to the triggered events and to the nontriggered events, 
but the privacy amplification is applied together, namely, 
 after the two reconciled keys are concatenated. The key rate 
$R^{\rm (both)}$ in this strategy is given by
\begin{eqnarray}
&&R^{\rm (both)}/q=
-Q^{\rm (t)}f(E^{\rm (t)})H_2(E^{\rm (t)})
\nonumber \\
&&-Q^{\rm (nt)}f(E^{\rm (nt)})H_2(E^{\rm (nt)})
+Q^{\rm (nt)}\min_x \{(1+r_0) x
\nonumber\\
&&
+(1+r_1) \xi(x)[1-H_2(\epsilon(x))]\}.
\label{eq:MGLLPboth}
\end{eqnarray} 
The final key rate is thus given by $R=\max \{R^{\rm (both)},R^{\rm (t)}\}$.

Next, we assume a channel model and show numerical examples of 
the key rate $R$ as a function of the distance $l$.
Let $\eta_{\rm c}=10^{-\alpha l/10}$ 
be the channel transmission, $\eta_{\rm B}$ be 
the quantum efficiency of Bob's
detectors, and $\eta\equiv \eta_{\rm c}\eta_{\rm B}$.
The background rate $p_{\rm d}$ of each detector is the combination of the
rates of the dark count and the stray light, which are assumed to happen
independently. For simplicity, we assume that both detectors have 
the same background rate.
$Q^{\rm (t)}_n$ is then given by
\begin{eqnarray}
Q^{\rm (t)}_n/p^{\rm (t)}_n&=&1-(1-\eta)^n(1-p_{\rm d})^2,
\label{eq:Yn}
\end{eqnarray}
and $Q^{\rm (t)}$ is calculated by taking summation.
Let $e_{\rm d}$ be the probability that a photon sent from Alice 
hits the erroneous detector, which is independent of the length
of the quantum channel. Then we have, after some calculation,
\begin{eqnarray}
&&2Q^{\rm (t)}_ne_n/p^{\rm (t)}_n=1-(1-\eta)^n(1-p_{\rm d})^2
\nonumber \\
&&-(1-p_{\rm d})[(1-\eta e_{\rm d})^n-(1-\eta+\eta e_{\rm d})^n],
\label{eq:en}
\end{eqnarray}
and $E^{\rm (t)}$ is calculated by taking summation.
$Q^{\rm (nt)}$ and $E^{\rm (nt)}$ are calculated similarly.

The values of the parameters are chosen as follows.
Alice may
use a non-degenerate PDC and obtain visible and
telecom-wavelength photons in mode ${\rm A}$ and ${\rm S}$,
respectively. Therefore, 
 we assume a typical silicon avalanche photodiode 
for D$_{\rm A}$, which has $d_{\rm A}=10^{-6}$ and (a)
$\eta_{\rm A}=0.5$. We also show the case
 with (b) $\eta_{\rm A}=0.1$ to see the dependence on $\eta_{\rm A}$.
The remaining parameters are borrowed from 
the experiment by Gobby {\it et al.} \cite{GYSapl04}, 
which are $\alpha=0.21$ [dB/km], $p_{\rm d}=8.5\times10^{-7}$, $\eta_{\rm
B}=0.045$, $e_{\rm d}=3.3$ [\%],  
and $f(E^{\rm (t)})=f(E^{\rm (nt)})=1.22$.
For each distance $l$, we have chosen the optimum
value $\mu_{\rm opt}$ for $\mu$ so that 
the key rate is highest, and the result is shown in Fig.~\ref{fig:graph} as
curves (a) and (b). The step at $\sim 130$ km, more pronounced on curve (b),
appears since the nontriggered events cease to contribute to the final
key at this distance. Beyond this distance,
the difference in $\eta_{\rm A}$
 causes a slightly low key generation rate for (b). 
We have also shown [curve (f)] the key rate for
the conventional analysis with $d_{\rm A}=0$ and $\eta_{\rm_A}=1$.
The remaining parameters are chosen to be the same.
In comparison to this key rate with $O(\eta_{\rm c}^2)$ dependence,
the key rates in our new protocol scale as $O(\eta_{\rm c})$,
 and the improvement reaches several orders of magnitude as the 
distance gets larger. Let us emphasize again that the two protocols 
use exactly the same experimental setup.

For comparison, we included key rates for schemes using WCP with decoy states [curves (d) and (e)] \cite{LMCprl05}. 
At shorter distances, the difference comes from that of the optimal mean photon number. 
%The present scheme achieves positive key rates at longer distances than the scheme using WCP with decoy states [curves (d) and (e)] \cite{LMCprl05}. $\mu_{\rm opt}$ becomes smaller than that of WCP. 
For example, $\mu_{\rm
opt}$ of (d) is 0.48 while that of (a) is 0.19. %at 10 km. 
This may be caused by the higher multi-photon rate of PDC, whose
 photon number distribution $p_n$ is thermal. However, the present
 scheme has a positive key gain 
up to almost the same distances as with an ideal photon source [curve
(c)]. The fact that no additional elements are needed in the PDC
setup to beat PNS attacks makes it a viable candidate for the practical QKD.
Recently Ma {\it et al.}~\cite{MFDCTL06} have shown that the achievable distance of WCP is further improved by two-way classical communication and post-processing. This interesting scheme also improves that of our scheme.

Finally, it is worth to discuss the feasibility of the present
scheme. As shown in Ref.~\cite{MSNI06}, high photon-pair generation from PDC ($\mu=0.9$)
using PPLN devices is possible in current technologies. 
The repetition rate of our scheme will be limited by that of D$_{\rm A}$, but it can
be improved by a high-repetition photon detection scheme \cite{NSI06}. Unlike WCP
schemes, the achievable distances with the single-photon source and the PDC
source depend on the coupling efficiency between the source and the single-mode
fiber.  
In the case of PDC from PPLN waveguide, we can
estimate the coupling efficiency of more than 80 \% in the current experiment
\cite{LDRYFT05}, which still leads to a longer
achievable distance than WCP schemes. The PDC source and the
single-photon source also suffer from
other losses in mode S such as ones at the polarization rotator, so further reduction of the losses is an important
subject in the future experimental studies.

In conclusion, we have proposed an efficient QKD protocol with PDC, which 
utilizes the events discarded in the conventional PDC protocol to
 derive tighter bounds on the rate and the QBER of the single-photon
 part. The only difference between the present and the conventional
 protocol is the classical data processing. We found that the key rate 
is significantly improved in the new protocol.

We thank R. Namiki and F. Takenaga for helpful discussions.
This work was supported by 21st Century COE Program by the Japan Society
for the Promotion of Science and a MEXT Grant-in-Aid for Young
Scientists (B) No. 17740265.

\end{document}